\documentstyle[emulateapj]{article}

\lefthead{Clayton et al.}
\righthead{UW Cen}

\begin{document}

\title{The Ever Changing Circumstellar Nebula Around UW Centauri}

\author{Geoffrey C. Clayton$^1$, F. Kerber$^2$, Karl D. Gordon$^1$, Warrick A. Lawson$^3$, 
Michael J. Wolff$^4$,
D.L. Pollacco$^5$ and E. Furlan$^6$}

\altaffiltext{1}{Department of Physics and Astronomy, Louisiana State 
University, Baton Rouge, LA 70803; 
Electronic mail: gclayton, gordon@fenway.phys.lsu.edu}

\altaffiltext{2}{ST-ECF, Karl-Schwarzschild-Strasse 2, D-85748 Garching bei M\"{u}nchen, 
Germany;
Electronic mail: fkerber@eso.org}

\altaffiltext{3}{School of Physics, University College, University of New South
Wales,\\ Australian Defence Force Academy, Canberra ACT 2600, 
Australia; Electronic mail: wal@ph.adfa.edu.au}

\altaffiltext{4}{Space Science Institute, Suite 294, 1234 Innovation Dr.,
Boulder, CO 80303-7814; Electronic mail: wolff@colorado.edu}

\altaffiltext{5}{Isaac Newton Group, Santa Cruz de La Palma,
                      Tenerife 38780, Canary Islands, Spain; 
Electronic mail: dlp@ing.iac.es}

\altaffiltext{6}{Institut f\"ur Astronomie, Universit\"at Innsbruck,
Technikerstr. 25, Innsbruck, Austria
Electronic mail: elise.furlan@uibk.ac.at}

\begin{abstract}
We present new images of the reflection nebula surrounding the
R Coronae Borealis Star, UW Cen.  This nebula, first 
detected in 1990, has changed its appearance significantly.  At the estimated
distance of UW Cen, this nebula is approximately 0.6 ly in radius so the nebula 
cannot have physically altered in only 8 years.  Instead, the morphology of the 
nebula appears to change as different parts are illuminated by light from the
central star modulated by shifting thick dust clouds near its surface. These dust clouds
form
and dissipate at irregular intervals causing the well-known declines in the 
R Coronae Borealis (RCB) stars. In this way, the central star acts like a lighthouse 
shining through holes in the dust clouds and lighting up different portions of
the nebula.
The existence of this nebula provides clues to the evolutionary history of RCB stars possibly linking 
them to the Planetary Nebulae and the final helium shell flash stars.
\end{abstract}


\section{Introduction}

R~Coronae Borealis (RCB)
stars are a small group of hydrogen-deficient carbon-rich 
supergiants which undergo spectacular declines ($\Delta V$ up 
to 8 magnitudes)
at irregular intervals apparently tied to their pulsation cycles
(Clayton 1996).  A cloud of 
carbon-rich dust forms; temporarily eclipses the 
photosphere, revealing a rich ``chromospheric'' emission spectrum;
then disperses, allowing the star to return to maximum light.  
The dust is thought to form in patches or puffs, not as a complete shell.  
Only when the dust condenses in the line of sight will a deep decline occur, although 
clouds may form during every pulsation cycle somewhere over the surface of the star.
Thus, the events are irregular and unpredictable.
Extensive ground- and space-based observations of R~CrB, RY~Sgr 
and V854~Cen have pointed to a unified empirical model of the 
RCB decline (see Clayton 1996 and references therein).  
A close connection between pulsational phase and the time of dust 
formation---seen in RY~Sgr and 
V854 Cen---implies that condensation occurs near the star (Pugach 1977; 
Lawson et al.\ 1992, 1999).  
Observed timescales for 
radiative acceleration of the dust, eclipse of the chromospheric region, and 
dispersal of the dust add further support to the ``near-star'' model.

There are two major evolutionary models for the origin of RCB stars: 
the double degenerate and the final helium shell flash (Iben, Tutukov, \& Yungelson 1996). 
The former involves the merger of two white dwarfs, and in the latter a 
white dwarf/evolved Planetary Nebula (PN) central star
is blown up to supergiant size by a final helium shell flash.
In the final flash model, there is a close relationship between RCB stars and
PN. The connection between RCB stars and PN has recently become stronger, since 
the central stars of three old PN (Sakurai's Object, V605 Aql and FG Sge; 
Duerbeck \& Bennetti 1996; Clayton \& De Marco 1997; Gonzalez et al. 1998; Kerber 1998; Kerber 
et al. 1999)
have had observed outbursts that transformed them from hot evolved central 
stars into cool giants with the spectral and dust formation properties of an RCB star. 
This establishes a possible connection between RCB stars, central stars
of PN and the final flash scenario.  
On the other hand, there is evidence that after the final flash a central star
acquires RCB characteristics only for a short time: V605~Aql was an RCB star
in 1921, but by 1986 it appeared to have a much hotter spectrum $T_{eff} \sim$ 50,000 K (Clayton 
\& De Marco 1997).
If most final flash stars evolve
so rapidly, this scenario for the formation of RCB stars may not be able
to yield the number of observed RCB stars.
Moreover, the recent claim
that white dwarf merging times might not be as long as previously thought,
has made the double degenerate scenario a valid and appealing
alternative to the final flash scenario for the formation of RCB stars 
(Iben et al. 1996).

So the presence or absence of nebulae around RCB stars is of great interest.
An old PN that is no longer ionized
could still be seen around a cool RCB star in 
starlight reflected from dust.  An 
example
of such a reflection nebula is the nebulosity observed around UW 
Cen (Pollacco et al. 1991).
This is the only cool RCB star known to have a visible nebula\footnotemark[1].
 \footnotetext[1]{V348 Sgr, a hot RCB star has an ionized nebula surrounding it (Pollacco, 
Tadhunter, \& Hill 1990)}
However, an extended shell-like structure around R CrB has been 
detected at 100 \micron~(Walker 1986). 
 
\section{Observations}

The lightcurve of UW Cen has been provided by the American Association 
of Variable Star Observers (AAVSO, Mattei 1999, personal communication). The two observational epochs
discussed here occurred during very deep declines of UW Cen when the star was $\sim$6 mag 
below maximum ($V_{max}\sim$9.5).
Figure 1a shows the V-band image of UW Cen obtained on 1990 May 16/17 with the ESO New 
Technology Telescope when the star was at V=16.5.  The description of instrument setup and 
reduction is contained in Pollacco et 
al. (1991). The image scale is 0.3\arcsec~pixel$^{-1}$ and the seeing was $\sim$1\arcsec.
We have obtained new BVR-band images using the 1 m 
and 2.5 m telescopes at 
Las Campanas Observatory. 
At the 1\,m Swope telescope, we used the SITE\,\#1, chip (2048 x 2048 pixels)
which gave a field of view of almost 24\arcmin~at a scale of
0\farcs69/pixel.
At the 2.5 m du Pont telescope, we employed the TEK 5 chip
(2048 x 2048 pixels) with 24 $\mu$m pixels
resulting in a scale of 0\farcs26/pixel.
In all observations, two or three exposures of  100 to 600 s duration,
depending on the filter used, were taken and co-added during reduction.
The intensities in the 1m telescope images were calibrated using observations of the PG1323-086 field 
which were obtained before and after the UW Cen images (Landolt 1992). The 2.5 m images were 
then calibrated using the 1 m data.
Only the V-band image from the 2.5 m telescope is shown in Figure 1b. 
 It was obtained 
on 12 February 1998 when UW Cen was at V=14.4 and the seeing was better than 1\arcsec.

\section{Modeling the Nebula}
Figure 1 shows how different the UW Cen nebula appears in images taken just 8 years apart.  
The nebula is very tenuous with a maximum surface brightness of $\sim$21-22 mag/$\sq^{\arcsec}$.
In 1990, the nebulosity is seen predominantly to the west of the star with clumps to the northeast, east and southwest. In 1998, there is shell-like nebulosity covering about 45\arcdeg~to the north of the star and very little elsewhere.  
The nebula has a diameter of approximately 15\arcsec~centered on  the star. The 
best estimate of the
 distance to UW Cen is 5.5 kpc (Lawson et al. 1990). The physical radius of the nebulosity is then
about 
6 x $10^{17}$ cm or 0.6 ly.  In order for material to flow from the star to the edge of the visible 
nebula in only 8 years, an outflow velocity of $\sim$24,000 km $s^{-1}$ would be necessary. Such a large 
velocity has never been seen in an RCB star. Outflow velocities of a few hundred kilometers per second are 
typically
seen in the circumstellar gas during declines (Clayton 1996). If this is a final flash shell, then 30-40 
km $s^{-1}$ is a 
more reasonable velocity (Guerrero \& Manchado 1996). Therefore, the apparent changes seen 
between 1990 and
 1998 were not caused by an actual change in the distribution of dust clouds in the nebula.  The likely 
explanation is that the dust nebula 
around UW Cen is being illuminated through shifting holes in a changing distribution of newly formed dust near the star.  
Since both 
images were obtained during deep declines we can surmise that dust had recently formed around UW Cen 
in both instances. To investigate this 
idea, we 
have produced a simple model of UW Cen and its circumstellar dust.

We employed the DIRTY radiative transfer
model (Gordon, Witt, \& Clayton 1999).  This model uses Monte Carlo
techniques to calculate the radiative transfer through arbitrary
distributions of dust viewed from any angle.  
The inputs to DIRTY are the stellar distribution, in this case one star, the dust
distribution, and the dust grain properties. 
We have modeled the dust distribution in the UW Cen nebula using two concentric shells.
The model inner shell lies near the star representing the new dust clouds which form during a decline 
episode.  This shell is optically thick, $\tau_V$=3,  but contains a number of
holes through which starlight shines to illuminate the outer shell.
The model outer shell is a 1\arcsec~thick homogeneous shell of dust located at a
distance of 6\farcs5 to 7\farcs5 from the star.  In between the two shells, there is a uniform
distribution of dust having a density which is a factor of ten below that of the
outer shell.  The optical depth  of the nebula not including the inner
optically thick shell is small, $\tau_V=$0.19.  The dust albedo and scattering phase
function are computed assuming amorphous carbon grains similar to those deduced for V348 Sgr 
(Hecht et al. 1998).  
The size distribution of the dust grains is a modified gamma function over the interval 0.03 - 0.10 
\micron. 
The resulting dust albedo in the V-band is
0.025 and the Henyey-Greenstein scattering phase function
asymmetry is 0.14.
Very good fits to the two images were produced by varying the number, size and optical depth of the 
holes in the inner shell.

Our model images are shown in
Figure~2.  Only a few holes in the inner shell are necessary to fit the 
observed nebula. The covering factors of the inner shell in these models are 87\% and 89\% for
the 1990 and 1998 images, respectively.  These numbers include a 5\% contribution from photons 
passing through the $\tau_V$=3 inner shell. However, the covering factor of the hemisphere facing 
away from the line of sight is not strongly constrained in this model.
It is possible to uncover a significant proportion of
the back of the star, $\sim$20\% of the surface area of the whole star, without seeing
significant variation in the model images. 
The dust behind the star contributes less to the scattered light because of the slight asymmetry in the 
scattering phase function. 
The model used here was chosen for simplicity. It is not unique. The observations could also be 
modeled using a clumpy outer shell and a smaller covering factor for the star.
More images obtained during declines of UW Cen are needed to fully map the nebula.
When its true morphology is determined, the UW Cen nebula can be compared to the final flash 
stars. In A30, for example, the PN is symmetric while its final flash shell is very clumpy 
(Guerrero \& Manchado 1986).

\section{Discussion}

The B-V colors of the UW Cen nebula are bluer than the star near the star and then progressively redder toward the outer edge. These colors are typical of those observed in other reflection nebulae.
Our model implies that the total optical depth from the star through the visible nebula
is $\sim$0.2. 
Assuming the amorphous carbon grains used for our model, the dust mass is $\sim$6 x $10^{-4}$
$M_{\sun}$.  If we assume a normal interstellar gas-to-dust ratio and solar hydrogen abundance for the nebula, then 
the total  mass is $\sim0.2~M_{\sun}$ (Bohlin, Savage, \& Drake 1978).  UW Cen is extremely hydrogen deficient and nothing is 
known about the abundances in the nebula so this estimate in very uncertain. 
\begin{figure*}[h]
\begin{center}
\plottwo{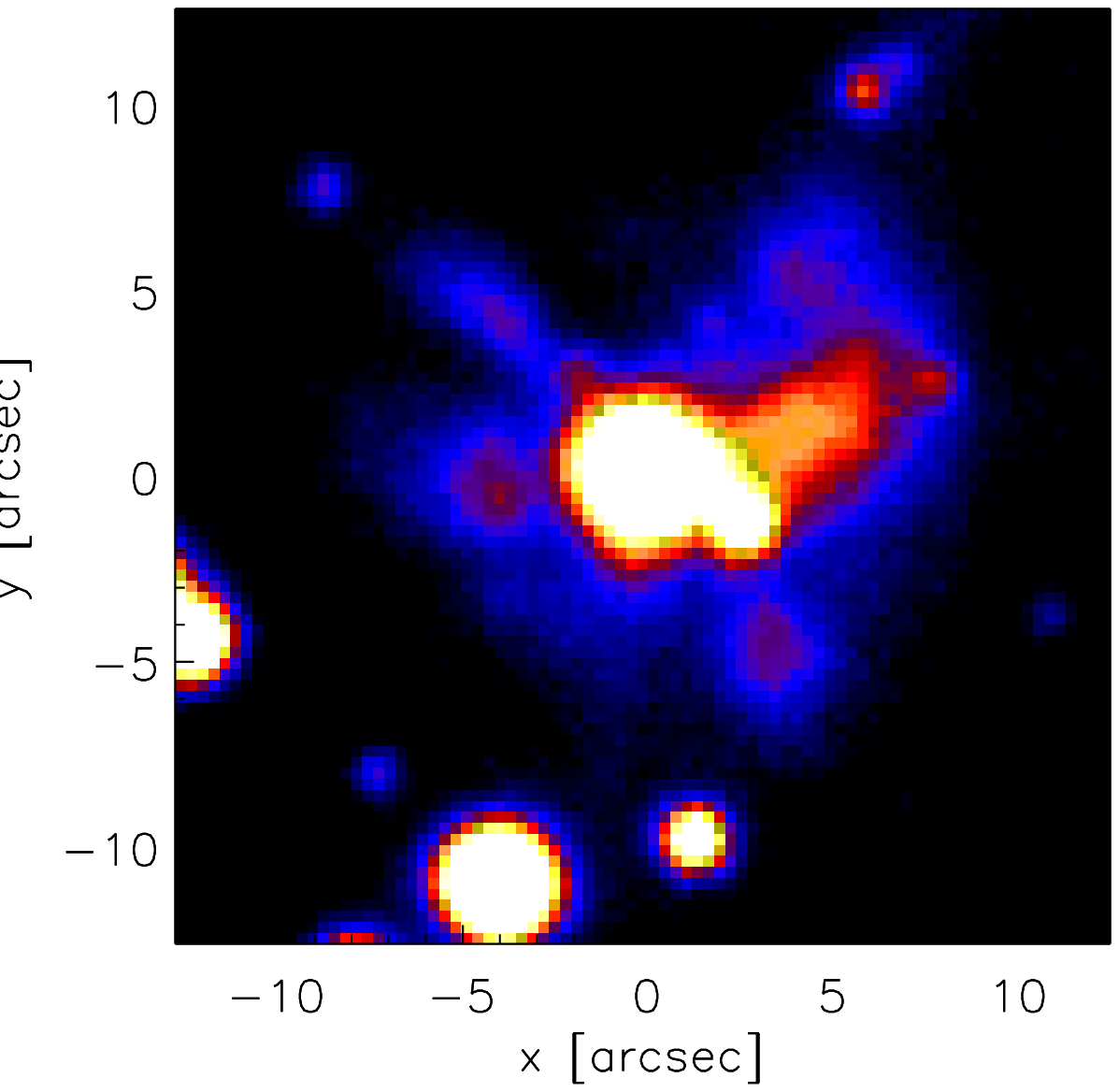}{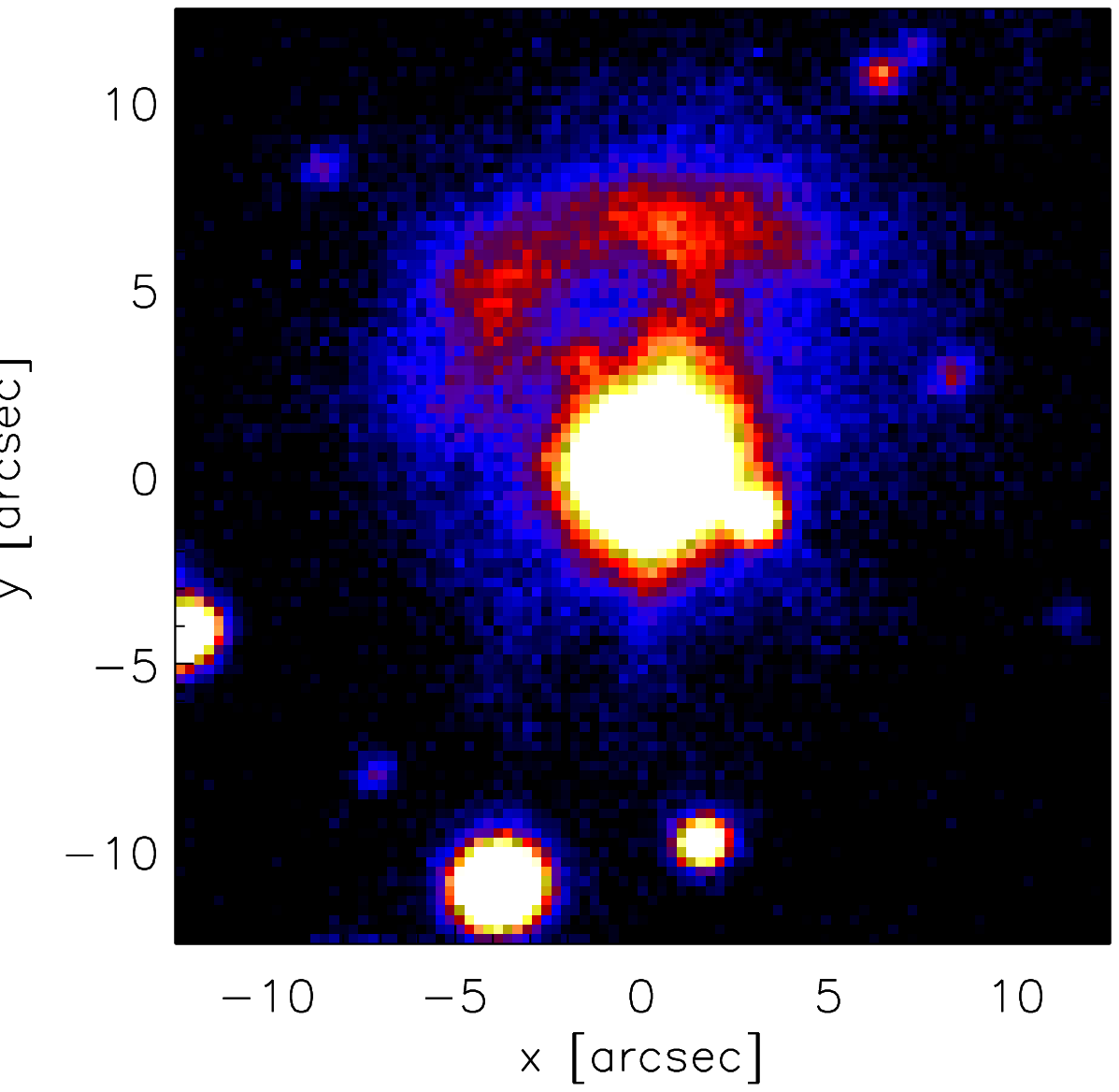}\\
\caption{(left) The UW Cen nebula in 1990 (Pollacco et al. 1991). 
(right) The UW Cen Nebula in 1998. Both images are plotted as log flux. 
The nebulosity ranges from 20.7 to 23.9 mag/$\sq^{\arcsec}$. 
The axes are RA offset (x) and declination offset (y). The sources 
which appear in Figure 1 but not Figure 2 are field stars.}
\plottwo{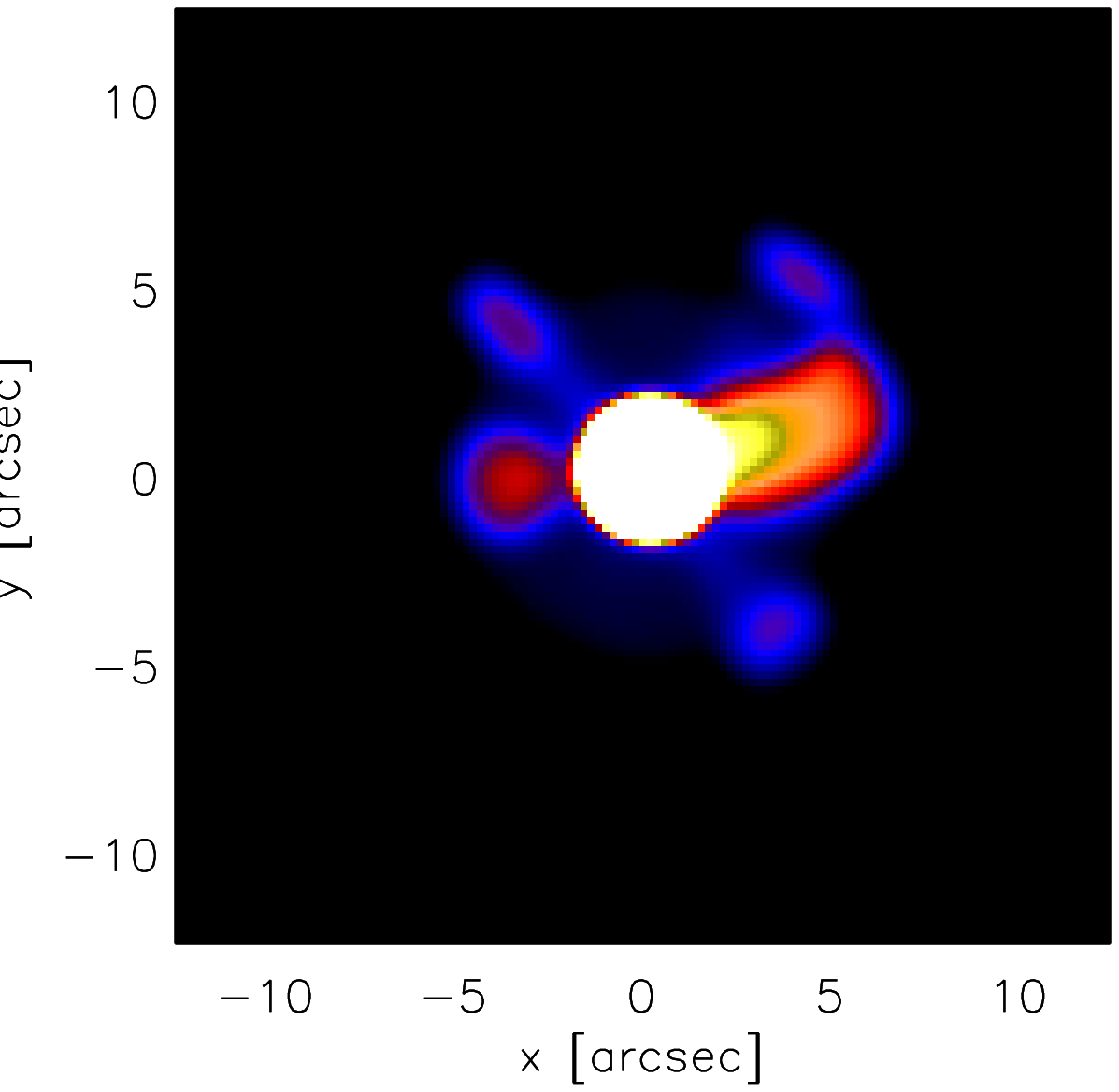}{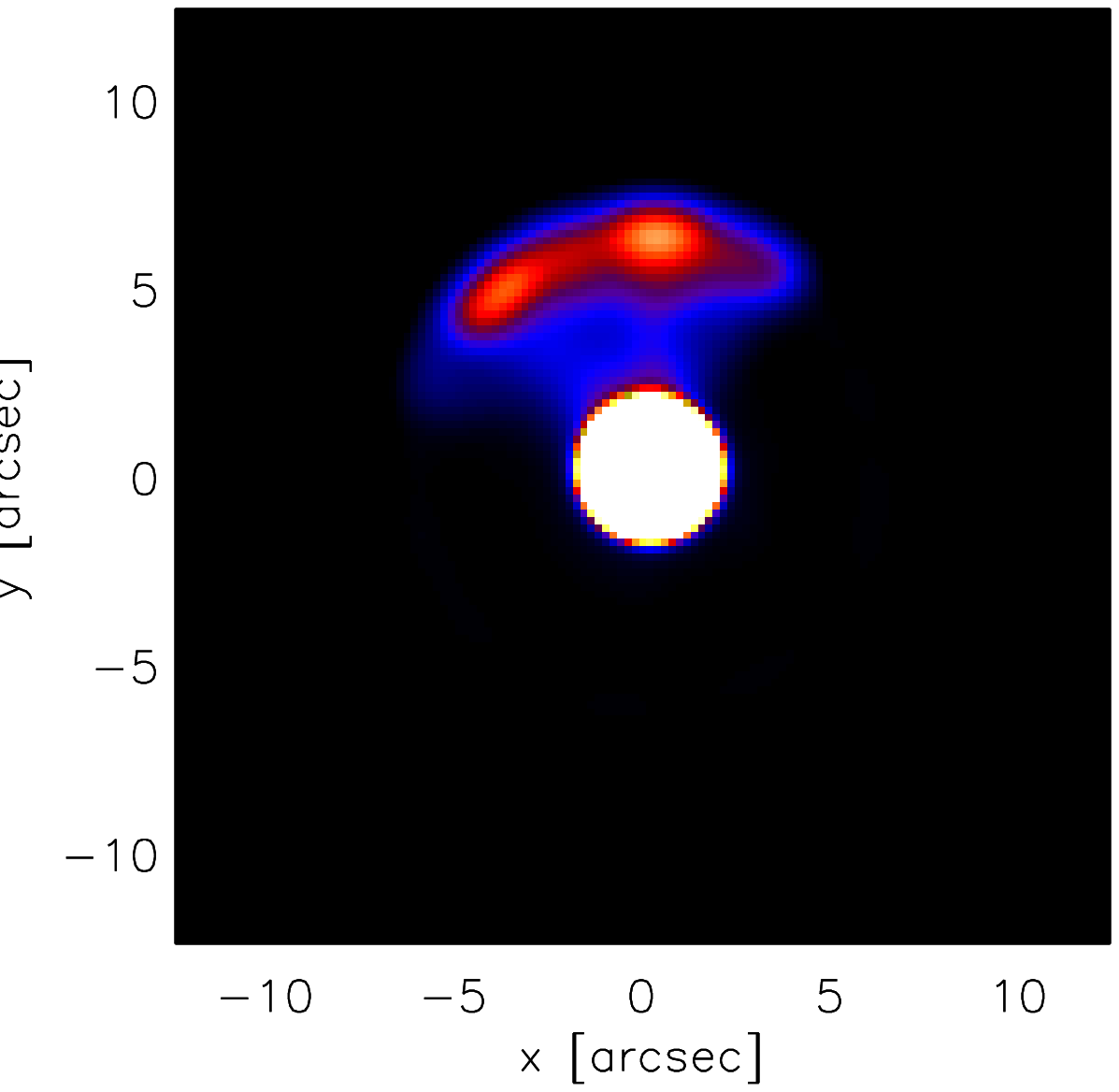}
\end{center}
\caption{Monte Carlo Dust Scattering Models. (left) The UW Cen 
nebula in 1990.  (right)  The UW Cen Nebula in 1998. They are plotted as 
described in Figure 1}
\end{figure*}

The mass of the final flash 
shell of A30 has been estimated to be $\sim$0.1 $M_{\sun}$.  The large IRAS shell around R CrB 
has an estimated mass of $\sim$0.3 $M_{\sun}$ (Gillett et al. 1986).

Comparisons of the direct emission and re-radiation from circumstellar dust in the IR give clues to the 
covering factor of new dust clouds forming around RCB stars (Clayton 1996 and references therein).
Figure 3 shows the flux distribution for UW~Cen.  The photometry of UW Cen were 
obtained from Jones et al. (1989), Kilkenny \& Whittet (1984) and 
the IRAS Point Source Catalog.  The optical and near-IR photometry
were obtained when the star was at, or near, maximum light.  These 
fluxes were dereddened assuming a small optical extinction ($A_{V} 
\approx$ 0.2 mag; Lawson et al. 1990) and using the typical diffuse interstellar 
extinction law (Rieke \& Lebofsky 1985; Cardelli, Clayton \& Mathis 1989). 

\vspace*{0.1in}
\begin{center}
\plotone{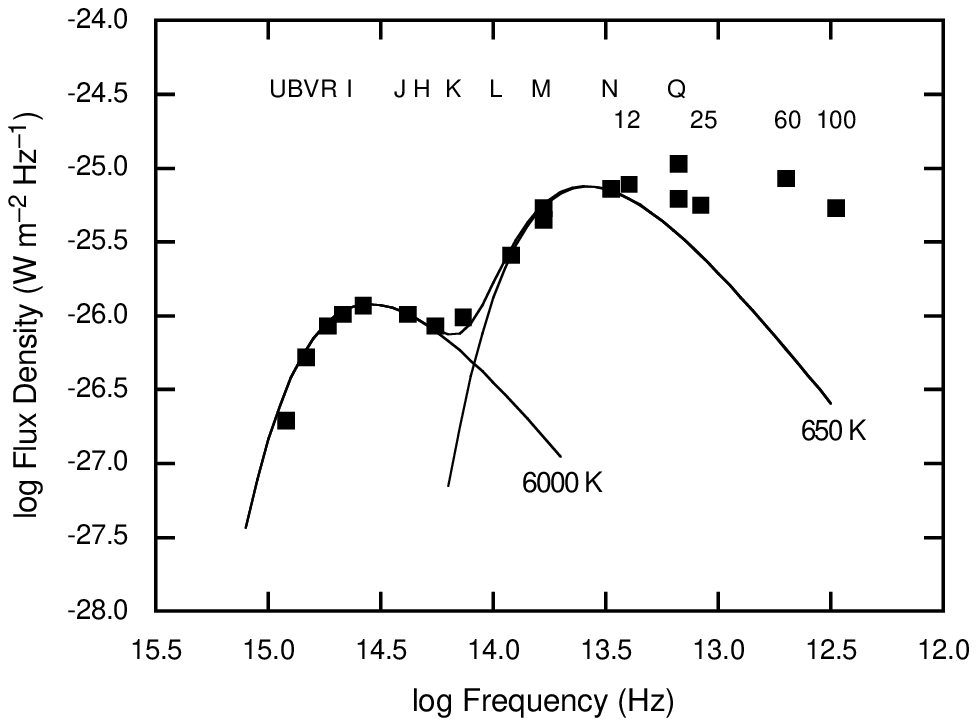}
\end{center}
\figcaption{The flux distribution of UW Cen.}

The stellar (optical and near-IR) flux and dust nebula flux (mid- and 
near-IR wavelengths) were fitted by two Planck functions with 
temperatures of 6000 $\pm$ 500 K and 650 $\pm$ 50 K, respectively.  
Planck functions tend to underestimate the $T_{\rm eff}$ of the 
star by 500--1000 K.  UW Cen is an F-type RCB star like R CrB and 
RY Sgr, which has $T_{\rm eff}$ of 7000 $\pm$ 500 K (Lawson et al. 1990). 
Wavelengths longer than 12 $\mu$m probably show the effect of 
bright IR cirrus towards UW Cen.  The cirrus contribution can 
be fitted with a 100 K Planck function.
The luminosity ratio {\it F}$_{\rm IR}$/{\it F}$_{\star}$ was 
calculated for UW Cen, where {\it F}$_{\rm IR}$ is the integrated 
flux of the Planck fit to the IR excess and {\it F}$_{\star}$ is 
that for the star.  For UW Cen, we derive a ratio of {\it F}$_{\rm IR}$/{\it F}$_{\star} \sim$ 0.3.
This is comparable to values given for UW Cen and other RCB stars based on long-term IR 
photometry 
(Feast et al. 1997). They find an apparent upper limit of {\it F}$_{\rm IR}$/{\it F}$_{\star}$
$\sim$ 0.5 for RCB stars. Since the albedo for carbon dust is so low, this ratio is a measure of the covering factor of the star. So typically, no more than 50\% of the stellar radiation is absorbed by the 
circumstellar clouds and re-radiated in the IR. 
No significant change is measured in the IR flux during an RCB decline indicating that the covering factor does not increase by a large amount during any dust formation episode.

Several studies of UV extinction due to the RCB circumstellar dust have estimated the covering factor (f) of the dust clouds by fitting the extinction and scattering in the dust.   
For the 
cool RCB stars, R CrB and RY Sgr, Hecht et al. (1984) find f$\lesssim$0.5. 
These estimates are
consistent with the visible/IR result and imply an upper limit of f=0.5 for RCB stars.  However, the results of our modeling of the UW Cen nebula imply a very high covering factor. 
This result is softened by the fact that 
the model is somewhat insensitive to the value of f on the far side of the star and that there are 
only data from two declines.  The covering factor undoubtedly varies from decline to decline.

If UW Cen is a final flash object then it will be an old PN central star and should be surrounded by the now neutral PN as well as the final flash shell. 
Seven PN in the Galaxy, A30, V605 Aql, A78, Sakurai's object, FG Sge,
IRAS 18333--2357 and IRAS 15154--5258 are hydrogen deficient and have central stars which have experienced a final helium shell flash (e.g., Guerrero \& Manchado 1996; Clayton \& De Marco 1997, Jacoby, De Marco \& Sawyer 
1998; Gonzalez et al. 1998). 
In each of these objects, the old PN surrounds a smaller final flash shell.
Predicted time scales 
indicate that the final flash will happen
$\sim$9000-18\,000 yr after the first ejection depending on mass (Bl\"{o}cker
1995, Iben et al. 1983) and 
that the 
star will rejoin the AGB a few hundred years after the final flash.
If we assume the star becomes an RCB shortly afterwards (as indicated by Sakurai's 
object), and stays
that way for $\sim$3000 to 10\,000~yr (Iben et al. 1996) and the dust is expanding at
35 km $s^{-1}$, 
then we would expect the old PN for UW Cen to have a diameter of 24\arcsec~to 50\arcsec, its
final flash shell to have a diameter of 8\arcsec~to 26\arcsec~and RCB dust inside that.  These diameters agree well with the measured 
value of 15\arcsec~for the UW Cen nebula.
Since only one nebula has been detected around UW Cen, it could be either an old PN or the final 
flash shell. 
Alternatively, the nebula could be a wind-blown bubble in the ISM. At 20 km $s^{-1}$, wind-blown 
dust could form the observed nebula in 9,500 yr.
A link between RCB stars and PN, if it can be established, will 
be an important step forward 
in understanding the evolution of post-AGB stars.
More observations are necessary to both map out the nebula of UW Cen and to resolve the 
disagreement between the large dust covering factor predicted by the model and the smaller values 
inferred from UV and IR data.

\acknowledgments
We thank the referee, Joel Kastner, for many useful suggestions.
We are grateful to Dr. Janet Mattei of the AAVSO for providing the lightcurve data for UW Cen. We 
thank Dr. O. De Marco for many helpful discussions.
It is a pleasure to thank Dr. M. Roth and the staff of Las Campanas Observatory
for their support. F.K and E.F acknowledge a travel grant from the Austrian
Ministry of Science and the University of Innsbruck.
Special thanks to Londo Mollari. This project was supported by NASA grant JPL 961526.


\end{document}